# Inventions on using sound and speech in Graphic User Interface
## A TRIZ based analysis

**Umakant Mishra**

Bangalore, India

http://umakantm.blogspot.in

**Contents**

## 1. Introduction

The personal computers are often equipped with a sound card that is capable of audio processing. This enables GUI applications to use sounds and synthesized speech, such as warning messages, voice alerts and guides to the user. Most sound cards also take input from a microphone, which are often used to pick-up user voice and digitize them through audio processing circuitry.

**Voice recognition system**

Voice recognition facilitates a human interaction with the machine. VR may be used to replace the manual task of pushing buttons on a wireless telephone keypad. This is useful when the hands of the user are busy with other activities like driving a car.

Voice recognition enables the user to enter information into the computer by speaking that information. For example, the voice recognition software can be used to enter text into a word processor program instead of typing the text on a keyboard.

Normally the voice recognition system needs to be trained before it can function optimally. The system should learn the speech patterns of a particular user. The digitized audio signals are stored with their text equivalents and compared with the voice being spoken.

In specific situations the voice recognition (VR) device should be capable of continuous speech recognition (CSR) to recognize a continuous speech, such as, speaking/ recognizing a telephone number or email address, which are generally spoken without a pause.

**Need and Application of voice enabled system**
- ⇨ The computer users may like to use VRS when their hands are busy. For example, a person can enter text through dictation while driving a car.

- ⇨ Voice enabled systems are useful for users having poor or partial vision. The voice commands supplement to the visual commands.

- ⇨ The VRS can be effectively used in maintaining phone books, address books etc. The user can dial telephones just by calling a telephone number or a name.

- ⇨ Sound is used in various applications to inform "alert", "error" or "ok" conditions.

- ⇨ The multimedia applications and gaming applications often use voice messages to say "sorry", "that's right", "thank you" etc. to make the session more interactive and attractive.

**Limitations of speech enabled system**
- ⇨ The VRS requires lot of training and customization to be effectively used by individual users as each individual falls into different voice patterns.

- ⇨ The voice interface is complex and error prone. Calling a telephone number does not give you as much certainty as pressing a key (or keys) for a telephone number.

- ⇨ Voice commands are generally associated with equivalent keyboard or mouse actuated control functions. They cannot operate commands on a background window.

- ⇛ Voice commands allow only a single command of a given control and not two or more commands of a given control.

- ⇛ VRS is more useful for visually impaired users, but there is less commercial viability in this market segment.

## 2. Inventions on speech enabled interface

There are many inventions on TTS, VRS and pattern matching techniques. We will not get into the speech mechanism rather discuss only on voice-enabled user interfaces. The following are six inventions selected from US patent database.

**2.1 Method and system for enabling a blind computer user to handle message boxes in a graphical user interface (US Patent 5223828)**

**Background problem**

Conventionally the message boxes are displayed in a box with options as "yes", "no" etc. This method is quite convenient for the normal user but does not help the visually impaired. There is a need to enable the visually impaired user to interact with such message boxes.

**Solution provided by the invention**

McKiel Jr. invented a method of enabling blind users to handle message boxes (US patent 5223828, Assigned by IBM, Jun 93). As per the invention, initially when the message box first appears, the text contents are audibly announced (using TTS or so). Then a homing signal is given by using a tone. The pitch of the homing sound changes according to the distance between the buttons and the mouse pointer. When the mouse pointer moves closer to the options there is an increase in the pitch of the homing sound which gives a feedback to the user whether the pointer is moving closer to or going away from the option buttons. When the pointer moves on to a push button, the caption of the push button is re-announced. Thus the user moves the mouse from button to button to hear the options and finally selects the one desired.

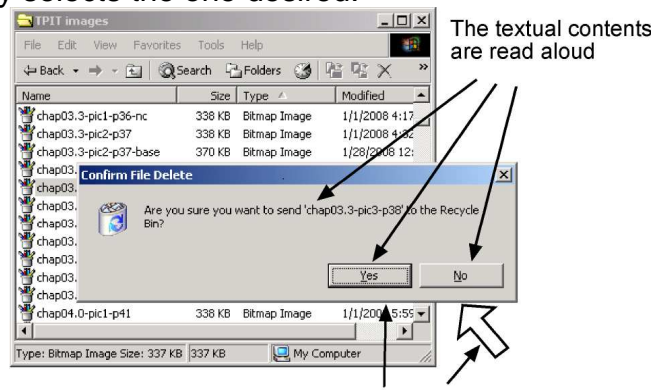

The textual contents are read aloud

The pitch of the sound increases as the distance between the pointer and message box decreases.

## TRIZ based analysis

The invention uses sound and speech to compensate the weakness of visual interface (Principle-8: Counterweight).

When the user moves the mouse pointer on to a push button the system audibly announces the caption of the button (Principle-23: Feedback).

The pitch of the sound increases when the distance of the mouse pointer and message box decreases and vice versa (Principle-35: Parameter change).

## 2.2 Method and system for enabling a blind computer user to locate icons in a graphical user interface (US Patent 5287102)

### Background problem

The previous invention although helps the blind user to handle a message box, it does not help her to locate other icons on the desktop. There is a need a mechanism to enable her to locate different icons on the desktop and visualize their positions on the screen.

### Solution provided by the invention

McKiel Jr. invented a method (US patent 5287102, assigned to IBM, Feb 94) of enabling a blind user to locate icons on a graphical user interface. According to the invention the system will provide audio information about the position of the pointer on the screen. When the pointer is located in the background, the system generates one type of sound and when the pointer is on an icon, it generates a different type of sound. When the pointer is positioned in the background it gives a stereophonic sound to inform about the location of the pointer whether positioned towards the left or right side of the screen. Besides, when the pointer is on an icon positioned in the left side there will be more sound from the left speaker and when the pointer is on an icon positioned in the right side there will be more sound from the right speaker. This stereophonic effect helps the user to know the position of icons on the screen.

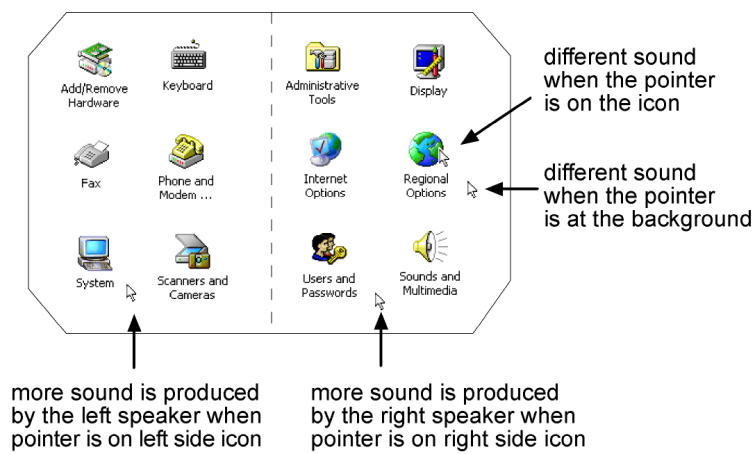

**TRIZ based analysis**

The invention uses sound and speech to compensate the weakness of visual interface (Principle-8: Counterweight).

The invention uses different types of sounds to indicate the position of the cursor; e.g., one type of sound when the pointer is positioned on an icon and different type of sound when on the background (Principle-35: Parameter change).

The method uses a stereophonic sound to inform about the location of the pointer whether positioned more towards left (by producing more sound from left speaker) or more towards right (by producing more sound from right speaker) side of the screen. (Principle-17: Another dimension).

**2.3 Method and system for enabling visually impaired computer users to graphically select displayed objects (US Patent 5461399)**

**Background problem**

In the previous invention the blind user knows about positions of different icons on the desktop. But in case of a GUI environment having multiple windows, the blind user cannot know whether the pointer is on window-1 or on window-2, or the pointer is on an icon in window-1 or is on an icon in window-2. There is a need to develop a method for the blind users to differentiate the icons in one window from another.

**Solution provided by the invention**

Brian J. Cragun disclosed a method of solving the above problem (US Patent 5461399, Assigned to IBM, Oct 95) by using sound. According to the invention, two different sounds will be associated with each object on the screen. The first sound is common among all objects in the class (i.e., same for all icons in a window) but unique among other classes (i.e., different for icons in different windows). The second sound is unique to the object in the class (i.e., different sounds for each individual icon in the window).

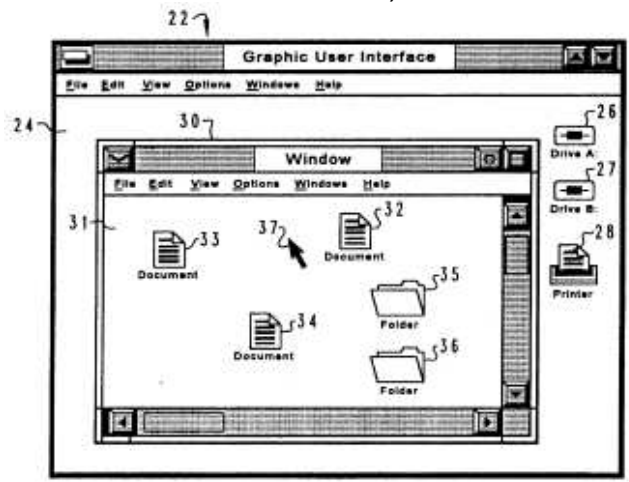

By listening to the audio signals of the objects the visually impaired user can identify different objects inside the class (i.e., within same window) and distinguish them from the objects in other classes (i.e. within other windows).

**TRIZ based analysis**

The invention uses sound and speech to compensate the weakness of visual interface (Principle-8: Counterweight).

The invention uses two different sounds, one indicating the window and other indicating the object inside the window. (Principle-35: Parameter change).

Both the sounds are used together to indicate a graphic object on the screen (Principle-5: Merging).

**2.4 Internal window object tree method for representing graphical user interface applications for speech navigation (US Patent 5864819)**

**Background problem**

Voice recognition systems respond to voice commands control software and allow user data to be entered through spoken words. These voice navigators are designed to associate voice commands with equivalent keyboard or mouse actuated control functions. However, the prior art voice navigators were not able to operate commands on a background window.

**Solution provided by the invention**

Patent 5864819 (invented by De Armas et al., assigned by IBM, issued Jan 1999) disclosed a voice navigator that can use the controls in a background window.

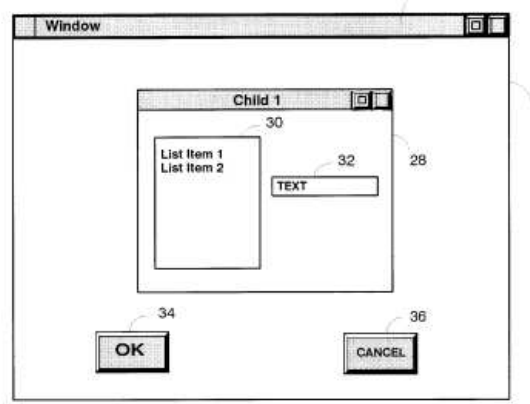

The invention enables an application program to determine a plurality of application states each defined as a set of window objects. The method internally maintains a window hierarchy by representing the application state by a sub-context tree comprised of various sub-context objects.

**TRIZ based analysis**

The invention differentiates a foreground state from a background state and maintains each state of the application as sub-context objects in context data file (Principle-24: Intermediary). The spoken words are matched with the vocabulary of the context data file.

**2.5 Method and system for enabling blind or visually impaired computer users to graphically select displayed elements (US Patent 6046722)**

**Background problem**
In a graphical user interface the objects are presented in visual icons, which the user operate through a pointing device like mouse. Normally the sighted individuals find graphical user interfaces very useful and convenient. But they are of no use for the blind or visually impaired people.

There are some methods to add sound and voice to the GUI so that they can be effectively used by the blind and visually impaired users. The audible cursor positioning and pixel status identification mechanism helps a user to identify the desired line by noting the change in tone as the cursor moves.

There is another system where the textual names of GUI contents are announced when they are first loaded. The blind user can select the graphical icons as their textual names are announced every time the mouse moves over them.

However these techniques do not help the user to select or deselect a displayed object. Hence, there is a need for an improved method for a blind or visually impaired user to graphically select the displayed elements within a graphical user interface.

**Solution provided by the invention**
Patent 6046722 (invented by Jr. McKiel, assigned by IBM Corporation, issued Apr 2000) provides a system for permitting blind or visually impaired users to utilize a graphical user interface.

According to the invention, a unique identifiable audible signal is associated with each displayed graphic element. A composite audible signal is periodically generated in response to the position of the cursor. The composite audible signal includes the audible signals associated with the graphic objects located within a predetermined radius from the cursor.

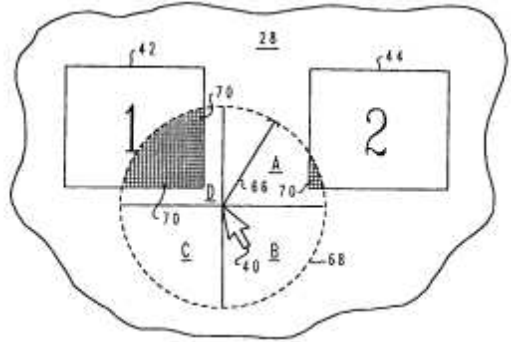

### TRIZ based analysis

The invention intends to provide a sound enabled system to assist the blind users to use the graphical user interface (Principle-8: Counterweight).

An audible signal is associated with each graphic element (Principle-38: Enrich).

The invention detects the position of the cursor and generates a composite audible signal that includes audible signals associated with the graphic elements within a predetermined radius (Principle-40: Composite).

## 2.6 Voice interaction method for a computer graphical user interface (US Patent 6499015)

### Background problem
A typical GUI displays icons, words and other graphical elements on the screen of a video monitor. This method is an improved and efficient alternative to typing commands through a keyboard. But this makes use of excessive and precise mouse movements, which may be difficult for some users. A voice enabled graphical user interface overcomes that difficulty.

### Solution provided by the invention
Patent 6499015 (invented by Brooks et al., assigned by IBM, issued Dec 2002) discloses a method of selecting elements of a GUI by speaking commands through a microphone.

According to the invention, the user speaks a voice command corresponding to the visual components such as menu items, command buttons etc. The computer then associates the voice command to the corresponding display element and does the necessary operation as if the command is entered through a mouse operation.

**TRIZ based analysis**

The voice recognition system supplements human computer interaction when the hands of the user are not available or busy with other activities (Principle-8: Counterweight).

When a voice command identifies an element on the GUI display the computer displays a menu of voice commands associated with that element (Principle-23: Feedback).

## 3. Summary

The TTS and VRS are not so popularly used with graphical user interfaces as it was expected a decade ago. This is mainly because the system is yet to be matured. The common user did not prefer such interfaces because of their complicacies. The application of TTS and VRS remained desirable only for specific segment of users.

But there are continuous inventions on the pattern matching and voice recognition mechanism. There will be more voice-enabled interfaces when the VRS mechanism will be more matured and fool proof.